\documentclass[10pt,twocolumn,letterpaper]{article}

\usepackage{cvpr}
\usepackage{times}
\usepackage{epsfig}
\usepackage{graphicx}
\usepackage{amsmath}
\usepackage{amssymb}
\usepackage{float}
\usepackage{xcolor}

\usepackage[breaklinks=true,bookmarks=false]{hyperref}

\cvprfinalcopy 

\def\cvprPaperID{****} 

\begin{document}

\title{Data Augmentation using Feature Generation for Volumetric Medical Images}

\author{
    Khushboo Mehra\\
	2576512\\
	Saarland University\\
	{\tt\small s8khmehr@stud.uni-saarland.de}
	\and
    Hassan Soliman\\
    2576774\\
    Saarland University\\
    {\tt\small s8hasoli@stud.uni-saarland.de}
    \and
    Soumya Ranjan Sahoo\\
    2576610\\
    Saarland University\\
    {\tt\small s8sosaho@stud.uni-saarland.de}
}

\maketitle

\begin{abstract}

    Medical image classification is one of the most critical problems in the image recognition area. One of the major challenges in this field is the scarcity of labelled training data. Additionally, there is often class imbalance in datasets as some cases are very rare to happen. As a result, accuracy in classification task is normally low.  Deep Learning models, in particular, show promising results on image segmentation and classification problems, but they require very large datasets for training. Therefore, there is a need to generate more of synthetic samples from the same distribution. Previous work has shown that feature generation is more efficient and leads to better performance than corresponding image generation \cite{xian_lorenz_schiele_akata_2018}. We apply this idea in the Medical Imaging domain. We use transfer learning to train a segmentation model for the small dataset for which gold-standard class annotations are available. We extracted the learnt features and use them to generate synthetic features conditioned on class labels, using Auxiliary Classifier GAN (ACGAN). We test the quality of the generated features in a downstream classification task for brain tumors according to their severity level. Experimental results show a promising result regarding the validity of these generated features and their overall contribution to balancing the data and improving the classification class-wise accuracy.
    
\end{abstract}


\section{Introduction}
 
    Brain tumor is considered as one of the deadliest and most common form of cancers. As a result, it is crucial to determine the correct type of brain tumor in early stages. It is of significant importance to devise a precise treatment plan and predict patient's response to the adopted treatment \cite{afshar_mohammadi_plataniotis_2018}. Accurate detection of these tumors from the surrounding healthy tissue and identifying its type is extremely important in the Medical field.
    \newline \par
    
    Detection of these tumors heavily depends on imaging techniques such as Magnetic Resonance Imaging (MRI). They are 3D images capturing the details of internal parts of human body. MRI a superior imaging modality in
    both lesion detection and in definition of tumor extent. MRI is better than or equal to other image acquisition techniques e.g. Computed Tomography (CT) in defining the anatomic extent of the tumor \cite{weekes_berquist_mcleod_zimmer_1985}. 
    \newline \par
    
    The specific type of brain tumor is determined by what the tumor looks like from the scanned images. Gliomas are the most common type of brain tumor in both children and adults. The word glioma encompasses many different tumor types, and gliomas come in different "Grades" \cite{research}. They are often categorized as Low Grade Gliomas (LGG), meaning that the tumor cells look as if they are dividing more slowly under the microscope, or High Grade Gliomas (HGG), HGGs appear as massively irregular regions and hence are easy to identify visually. On the contrary, LGGs have vague and smooth boundaries and can be challenging to be detected even with purely scanned MRI images.
    \newline \par
    
    It is argued that image feature generation has many advantages over image generation \cite{xian_lorenz_schiele_akata_2018}. The number of generated image features has no bounds. Additionally, The image feature generation learns from compact invariant representations obtained by a deep network “U-Net” in our case, which is trained on a large-scale dataset of brain tumor images “BraTS”. Therefore, the generative network can be quite shallow and hence computationally efficient. Also these generated CNN features are highly discriminative. 
    \newline \par
    
    With this background, the main idea discussed here is based on the hypothesis mentioned above regarding the preference of image feature generation over image generation. We tested this hypothesis in the Medial Field by investigating brain tumor images. We examined its performance in classifying the type of these tumors by training a classifier on the dataset of augmented image features and checking the class-wise accuracy.


\section{Related Work}
   
   In summary, the work presented in this report builds on previous research to generate synthetic images for the purpose of data augmentation, especially in the field of medical imaging. In this section we review some recent relevant literature on Medical image data augmentation and Feature generation networks.
   \newline \par
   
   The authors in \cite{shin_tenenholtz_2018} used GANs to generate synthetic images for the purpose of data augmentation. Further, the authors in  \cite{dong_yang_liu_mo_guo_2017} have used a U-net based CNN to generate high quality segmented images. Also, we emphasize on the work in \cite{xian_lorenz_schiele_akata_2018}, where the experiments shows that image features perform better than the visual images. Note that although the work in \cite{xian_lorenz_schiele_akata_2018} focuses on zero-shot learning, our hypothesis is that their methodology would benefit datasets with class imbalance.
   \newline \par
   
   Therefore, in our work, we propose to tackle the task of generating class conditioned training samples for addressing the problem of data diversity in the medical domain, by synthesizing image features  conditioned on class labels. Hence, we consider our work to be a novel application in the medical imaging domain.


\section{Method}

    We propose the following method to answer the following questions:
    \begin{itemize}
        \item Can the pre-trained "U-Net" model be used to extract discriminitive features about the tumors types? 
        \item Can we generate more features from features extracted from the "U-Net", and use them to improve  performance on a downstream classification task?
    \end{itemize}


    \subsection{U-Net FCNN Architecture}
    
\begin{figure*}[h]
        \centering
        \includegraphics[width=0.75\textwidth]{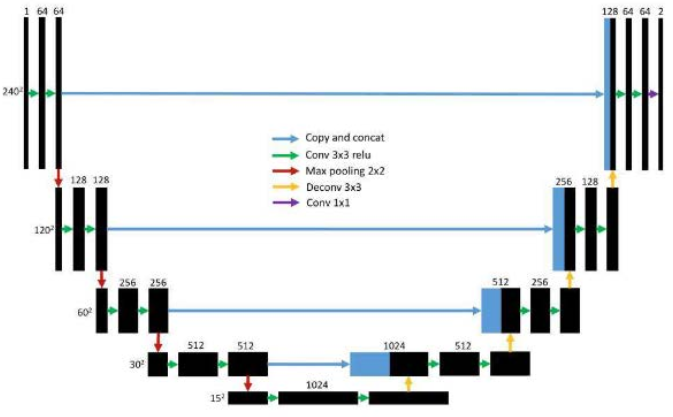}
        \caption{\label{fig:U-Net}The U-Net Architecture}
        \end{figure*}

        We use the U-Net model \cite{ronneberger_2017} pre-trained on BraTS dataset to perform the segmentation task. Fig. \ref{fig:U-Net} shows the U-Net architecture used for the BraTS segmentation challenge \cite{dong_yang_liu_mo_guo_2017}. The modified pre-trained U-Net we used consists of 4 forward encoding blocks and max pooling layer after each block. The encoding block takes an input image of size (240x240x4) and generates a (15x15x1024) sized encoding. Similarly, there are 4 decoding blocks with an upsampling layer after each deconvolutional operation. Moreover, these deconvolution layers also take as input the shallow representations from similarly sized layers in the encoding part of the U-Net. Finally, the U-Net outputs a pixel map of size (240x240x1) contains a binary mask of the segmented tumor as shown in Fig. \ref{fig:Original-and-Segmented-Tumor}. There is an open source implementation of the "U-Net" model \cite{zsdonghao_2019} in Tensorlayer \cite{dong_supratak_mai_liu_oehmichen_yu_guo_2017} and the pre-trained version of this model on BraTS is provided by \cite{amr_amer_ishwar_2018}, which we have re-used in our project. 
        \newline \par
        
        We further fine-tuned this pre-trained model on REMBRANDT dataset, so that it can learn more features about brain tumors from another similar dataset to BraTS.
        
    \subsection{Auxiliary Classifier GAN Architecture}
\begin{figure}[h]
        \centering
        \includegraphics[width=0.3\textwidth]{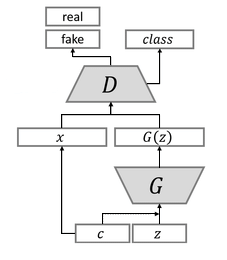}
        \caption{\label{fig:acgan}The AC-GAN Architecture: $G$ is the generator network that takes noise and class label, $D$ is the discriminator that predicts image source and class label.}
        \end{figure}

    For generating synthetic features, we use a variant of class-conditional generative adversarial networks, the auxiliary classifier GANs \cite{odena2017conditional}. Like a vanilla-GAN, the AC-GAN consists of two neural networks trained in opposition to one another. The generator $G$ takes random noise vector $z$ concatenated with the class label $c$ and generates a synthetic image feature $X_{fake} = G(c,z)$. The discriminator network $D$ receives either a real feature or a fake feature, and outputs both a probability distribution over sources (real vs fake) and a probability distribution over the class labels, $P(S|X), P(C|X) = D(X)$ (Fig. \ref{fig:acgan}). The objective function is composed of two parts: log-likelihood of correct source $L_S$ and log-likelihood of correct class $L_C$.
    \begin{equation}
        L_S = E[log P(S = real | X_{real}] + E[log P(S = fake | X_{fake}]
    \end{equation}
    \begin{equation}
        L_C = E[log P(C = c | X_{real}] + E[log P(C = c | X_{fake}]
    \end{equation}
     The discriminator is trained to maximise $L_S + L_C$, and the generator is trained to maximise $L_C - L_S$. The authors show that the modified objective stabilizes training and results in higher quality of generated samples compared to the vanilla-GAN.
     \newline \par
     $D$ and $G$ use architectures with convolutional and deconvolutional layers, respectively. On the other hand, \cite{xian_lorenz_schiele_akata_2018} use MLP based architectures for the generator and discriminator models because the image features are already extracted using a CNN. However, our work uses architecture similar to \cite{odena2017conditional} because the features extracted from the U-Net have a very high dimensionality (15x15x1024) compared to \cite{xian_lorenz_schiele_akata_2018} (2048). We found that downsampling results in severe information loss (Sec. 4.1.4). Therefore, using a CNN based architecture with original features was more memory efficient. There are two differences between our implementation and the model from \cite{xian_lorenz_schiele_akata_2018}. First, we decreased the number of hidden layers in both networks from 5 to 3 because our feature maps are much smaller (15x15) compared to normal images that the AC-GAN was originally designed to process. Second, we changed the number of feature maps the hidden layers produce to suit our data which has 1024 channels compared to standard images, which have 3 channels. Both networks use leaky relu activation. We followed a publicly available ACGAN implementation in pytorch \footnote{https://github.com/gitlimlab/ACGAN-PyTorch} and changed the architecture as mentioned above, along with other changes to make the code compatible with the latest version of pytorch.
     
    \subsection{Binary Classifier}
    \cite{yi2018generative} states that a major shortcoming of most GAN based data-augmentation approaches in medical imaging is that they do not evaluate performance on a different, downstream task. Hence, we propose to test the quality of the synthetic features by classifying the data into 2 classes: HGG and LGG. While using a linear classifier is more straightforward, we use a CNN based classifier for high-dimensionality issues already covered in the previous section. 
    
    \subsection{Dataset}
    
\begin{figure}[h]
        \centering
        \includegraphics[width=0.5\textwidth]{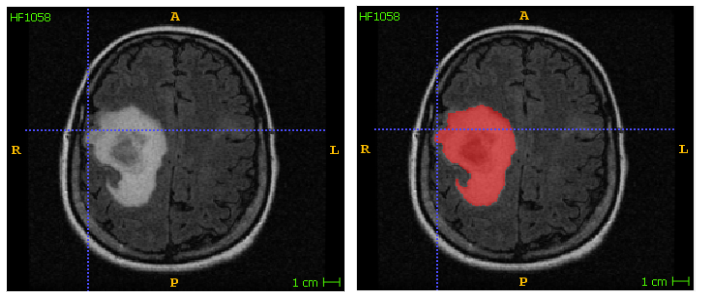}
        \caption{\label{fig:Original-and-Segmented-Tumor}Left: Example Extracted Slice for a Sample for HGG Patient. Right: Corresponding Segmented Slice}
        \end{figure}

        The REMBRANDT dataset \cite{scarpace2015data} contains multimodal MRI Brain Tumor images of Glioma patients. We focused only on one type of modality called “FLAIR” which consists of a certain type of MR pulse sequences that can capture the whole region of the tumor as shown in Fig. \ref{fig:Original-and-Segmented-Tumor}. In this figure, it shows one slice of a sequence of a random sample from REMBRANDT dataset, and the corresponding segmented tumor. They are 3D images with different number of channels each.
        \newline \par
        
\begin{figure}[h]
        \centering
        \includegraphics[width=0.5\textwidth]{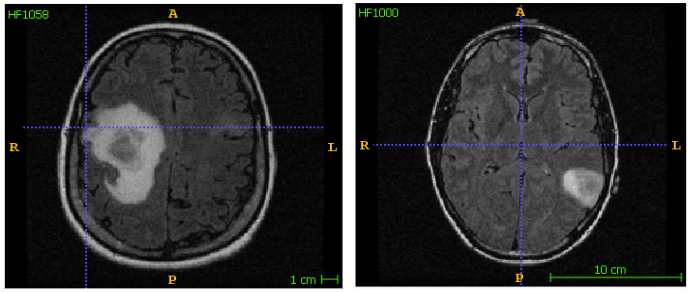}
        \caption{\label{fig:HGG-and-LGG-Tumor}Left: Example Sample for HGG Patient. Right: Example Sample for LGG Patient}
        \end{figure}

        We received this data for 101 patients. The dataset is divided and labeled according to the severity of tumor into “HGG” and “LGG” explained previously. As shown in Fig. \ref{fig:HGG-and-LGG-Tumor}, there is a challenge with “LGG” samples more than “HGG” samples as they are smaller and hard to be distinguished.
        
    \subsection{Data Preprocessing}
    
        Since the dataset is very small “101” cases, divided into (1) 61 cases labeled as HGG, and (2) 40 cases labeled as LGG, therefore 90\%-10\% division is recommended in this case resulting into (1) 54 training samples and 7 test samples for HGG, and (2) 36 training samples and 4 test samples for LGG.
        \newline \par
        
        As previously mentioned in (Sec. 3.4), each case from REMBRANDT dataset has different number of channels, so every channel from each case is extracted, now every channel is considered a sample in itself. After extraction, the resulted number of samples of training set is 4540 divided into (1) 2712 samples labeled as HGG, and (2) 1828 samples labeled as LGG, while the resultant number of samples of test set is 570 divided into (1) 362 samples labeled as HGG, and (2) 208 samples labeled as LGG.
        \newline \par
        
        Each sample extracted is of size (240x240x1), it's then normalized by subtracting the mean of all the samples, then divide by their standard deviation. Each corresponding sample from the target segmented samples has the same size of (240x240x1). 
    
    
    \subsection{Dataset Augmentation}
    
        Large dataset is important for any neural network model to be properly trained, even the "state-of-the-art" deep networks need massive amount of data to give promising results. Since REMBRANDT dataset is very small, Dataset Augmentation is crucial in this case. We experimented with using different techniques and came up with metrics as shown in Tab. \ref{table:1}. 
        \newline \par
        
        \begin{table}[h!]
        \centering
        \begin{tabular}{|p{3.0cm}| p{4.0cm}| }
        \hline
         Technique & Parameters \\
        \hline
        1. Flip Horizontally & 50\% Probability \\ 
        2. Flip Vertically & 50\% Probability \\  
        3. Elastic Distortion & $\alpha$ = 720, $\sigma$ = 24 \\
        4. Rotation & $\pm$ 20$^{\circ}$ \\
        5. Shift & 10\% Horizontally and 10\% Vertically \\
        6. Shear & 5\% Horizontally\\
        7. Zoom & $\pm$ 10\%\\
        
        \hline
        \end{tabular} \\
        \caption{Data Augmentation Techniques}
        \label{table:1}
        \end{table}
        
\begin{figure}[h]
        \centering
        \includegraphics[width=0.45\textwidth]{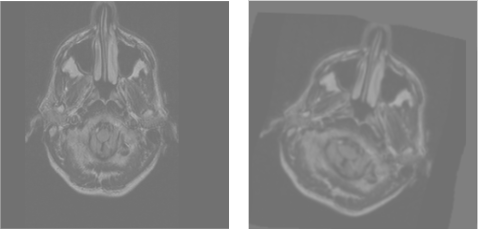}
        \caption{\label{fig:Original-vs-Augmented-Sample}Left: Example Original Sample of Training Set. Right: Corresponding Augmented Sample}
        \end{figure}

        We applied "Elastic Distortion" technique as tumors have no definite shape. In addition, we applied very slight "Shear" technique in the horizontal direction which doesn't affect the global shape of the tumor. All these methods contribute to generating more training data with slight variations compared to the original ones as shown in Fig. \ref{fig:Original-vs-Augmented-Sample} 
        \newline \par
        
\section{Experiments}

    We discuss the experiments conducted regarding the evaluation of the model, the training parameters used, and the final results for both models (1) U-Net Fully Convolutional Network, and (2) Auxiliary Classifier GAN. 
    
    
    \subsection{Feature Extraction}
    
        Using U-Net FCNN \cite{ronneberger_2017} shows that it can be trained end-to-end using very few images and outperforms the prior state-of-the-art model for segmentation problem. As previously mentioned, our hypothesis is that we can get better feature vectors of brain tumors using this network, which is already pre-trained on semantic segmentation of brain tumors on the "BraTS" dataset. We believe that this model is already learnt to extract the features of the fine details of the brain tumor, we further fine-tuned it on "REMBRANDT" dataset which is similar to "BraTS" dataset, so that it can learn more features about brain tumors that will help with the problem of classification and detection of its severity level (HGG or LGG).
    
        
        \subsubsection{Model Input Data}
            
            As previously mentioned, we divided "REMBRANDT" dataset into training set and test set according to 90\%-10\% ratio per each class. We used test set as a validation set to evaluate the model for each epoch. This dataset is preprocessed and augmented using methods mentioned in (Sec. 3.5 and 3.6).
            \newline \par
            
            Model weights are initialized from the saved weights of the pre-trained U-Net model on "BraTS" dataset. Our objective is to fine-tune it further on REMBRANDT dataset, then after fine-tuning, we extract the image features of the dataset, this will serve as input for the Auxiliary Classifier GAN (Sec. 4.2).

        
        \subsubsection{Model Hyperparameters}
        
            We trained the model for 30 epochs at a learning rate of 0.00001, and batch size of 10. Model is trained and optimized using Adam Optimizer with Soft-dice loss as a loss function. We used a Soft-dice based loss function explained in \cite{milletari_navab_ahmadi_2016} as shown in
            eq. (\ref{soft_dice_loss}). The Soft Dice based loss function has a unique advantage that is robust to unbalanced dataset, which is very important for brain tumor segmentation because some sub-tumoral regions may only count for a small portion of the whole tumoral volume \cite{dong_yang_liu_mo_guo_2017}. In addition, in the medical field, it's common to evaluate the segmentation based on Soft-dice metric instead of the well known Intersection over Union metric.  
        
        
        \subsubsection{Model Performance Evaluation}
        
            We evaluated the model against 3 metrics as follows, (1) Soft-dice loss as shown in
            eq. (\ref{soft_dice_loss}) , (2) Hard-dice score as shown in eq. (\ref{hard_dice}) , and (3) Intersection over Union as shown in eq. (\ref{IOU}). Soft-dice is a modified version of Hard-dice but with adding a very small number $\epsilon$ to both numerator and denominator for numerical stability.

            \begin{equation}\label{soft_dice_loss}
	            Soft Dice Loss = 1 - \frac{2TP + \epsilon}{(TP + FP) + (TP + FN) + \epsilon}
            \end{equation}
            \begin{equation}\label{hard_dice}
	            Hard Dice Score = \frac{2TP}{(TP + FP) + (TP + FN)}
            \end{equation}
            \begin{equation}\label{IOU}
	            IOU = \frac{TP}{(TP + FP + FN)}
            \end{equation}
        
        
        \subsubsection{Model Final Results}
        
\begin{figure}[h]
        \centering
        \includegraphics[width=0.5\textwidth]{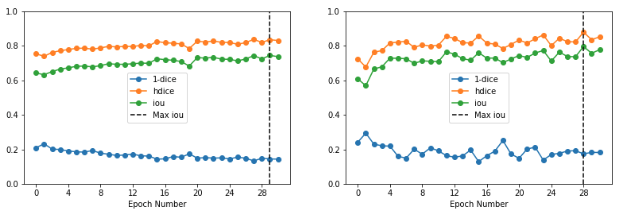}
        \caption{\label{fig:Training-and-Validation-Set-Metrics}Left: Training Set Evaluation Metrics. Right: Test Set Evaluation Metrics}
        \end{figure}

        As shown in Fig. \ref{fig:Training-and-Validation-Set-Metrics}, For the training set metrics, model is properly trained as Soft-dice loss is steadily decreasing, while Hard-dice score and IOU are steadily increasing. For the validation set metrics, the evaluation metrics of the model are slightly fluctuating but the final metrics are similar to those of the training set. Best values achieved for every metric for both training and validation sets are shown in Tab. \ref{table:2}. The black vertical line in Fig. \ref{fig:Training-and-Validation-Set-Metrics} is drawn at the epoch number where it recorded the best value for IOU.
        \newline \par
        
        \begin{table}[h!]
        \centering
        \begin{tabular}{|p{2.0cm}| p{1.5cm}| p{1.5cm}| p{1.5cm}|}
        \hline
         & Soft-dice Loss & Hard Dice Score & IOU \\
        \hline
        Training Set & 0.13 & 0.84 & 0.74\\ 
        Validation Set & 0.13 & 0.88 & 0.79\\  
        
        \hline
        \end{tabular} \\
        \caption{Data Augmentation Techniques}
        \label{table:2}
        \end{table}
        
\begin{figure}[h]
        \centering
        \includegraphics[width=0.45\textwidth]{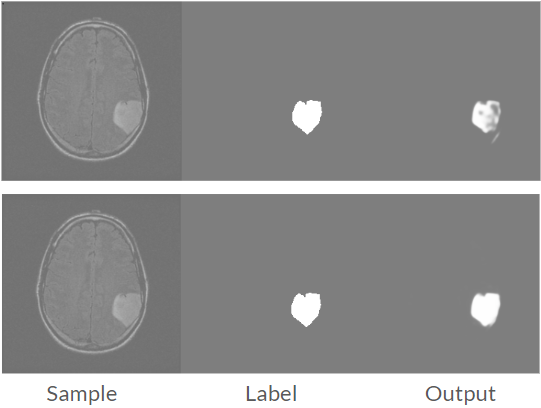}
        \caption{\label{fig:Fine-tuned-vs-Pre-trained-Model-Labelled}Top: Sample evaluated on the pre-trained model on "BraTS" dataset. Bottom: Sample evaluated on the fine-tuned model on "REMBRANDT" dataset}
        \end{figure}

        The fine-tuned model demonstrated a good improvement over the pre-trained model regarding solving the semantic segmentation problem of brain tumors as expected. Fig. \ref{fig:Fine-tuned-vs-Pre-trained-Model-Labelled} shows a comparison of the output of one sample, when evaluated against the pre-trained model on “BraTS” dataset, and when evaluated against the same model fine-tuned on “REMBRANDT” dataset.
        \newline \par
        
\begin{figure*}[h]
        \centering
        \includegraphics[width=1\textwidth]{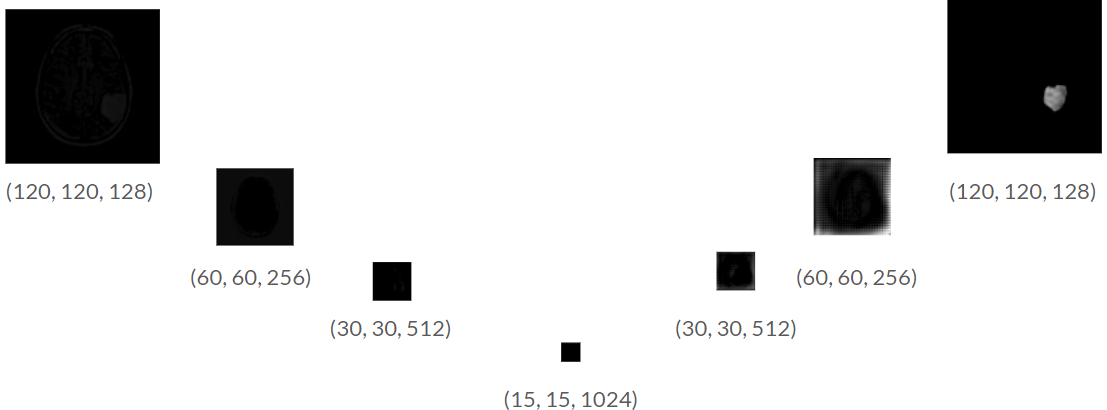}
        \caption{\label{fig:Features-Extracted}Resulted Features extracted at each block of the U-Net model}
        \end{figure*}

        Fig. \ref{fig:Features-Extracted} shows an example of features extracted from different blocks of the U-Net, for the sample in Fig. \ref{fig:Fine-tuned-vs-Pre-trained-Model-Labelled}. It also shows the corresponding feature size we get from each block. We ran the fine-tuned model against the whole "REMBRANDT" dataset, and extracted the features from the output of the 5th convolutional block in the "U-Net", resulting into features of size (15x15x1024).
        \newline \par
        
        Although it is better to use the features extracted from the higher layers in the expansion part of the "U-Net" e.g. features generated from the 3rd deconvolutional block of size (120x120x128) as shown in Fig. \ref{fig:Features-Extracted}. We faced a memory overflow issues with the RAM while loading the extracted features of this size (120x120x128) for all the samples of the dataset (n = 5110). We tried the same thing but with the extracted features of size (30x30x512), but RAM also crashes, it only worked with extracted features of size (15x15x1024). However, this means that we lose a lot of information about the images of the brain tumor because of downsampling. 
        
    
    \subsection{Feature Generation}
    
        
        \subsubsection{Model Input Data}
\begin{figure}[h]
        \centering
        \includegraphics[width=0.5\textwidth]{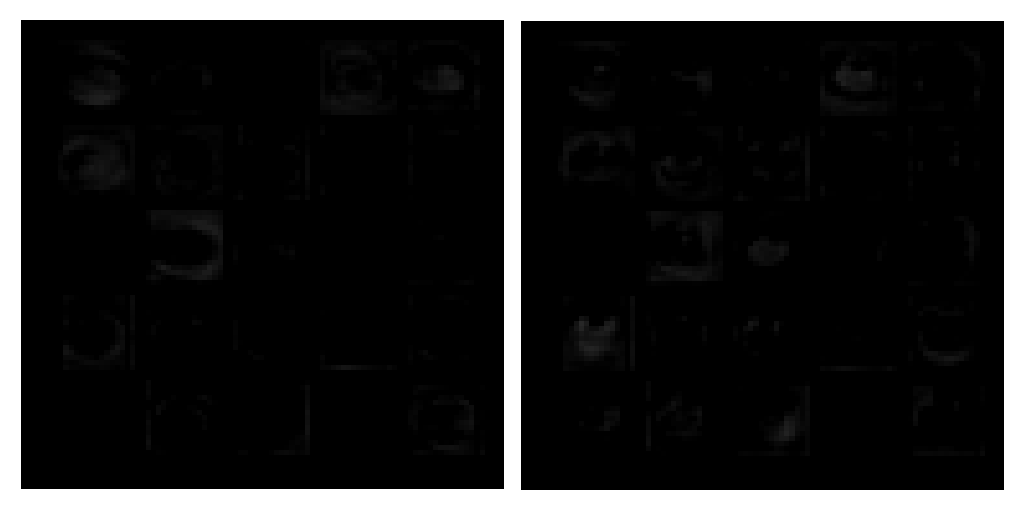}
        \caption{\label{fig:ft-sample}A sample of 25 feature maps from LGG (left) and HGG (right).}
        \end{figure}

        We use features of size (1024x15x15). While features from higher deconvolutional blocks in the U-Net would have more information, we could not use these to train because of memory limitation. The features are split into training set of 4540 features and a test set of 570 (reserved for testing the binary classifier). The generator takes random noise and class labels as inputs and produces synthetic features. The discriminator is trained on extracted and synthetic features. Samples of extracted features from both classes are shown in Fig. \ref{fig:ft-sample}.
        
        \subsubsection{Model Hyperparameters}
        The ACGAN was trained using learning rate of 0.002, batch size of 16 for 50 epochs. The discriminator uses binary cross entropy loss for image source and negative log-likelihood for class labels. Both networks use the Adam optimizer. We experimented by varying the number of channels $ndf$ and $ngf$ in the hidden layers of $D$ and $G$, respectively. We experimented by setting these hyperparameters to 64, 256 and 512. Using a low number of kernels would force the network to downsample while using large number of feature maps would retain more information and result in a better quality of generated features.
        
        \subsubsection{Model Performance Evaluation}
\begin{figure}[h]
        \centering
        \includegraphics[width=0.5\textwidth]{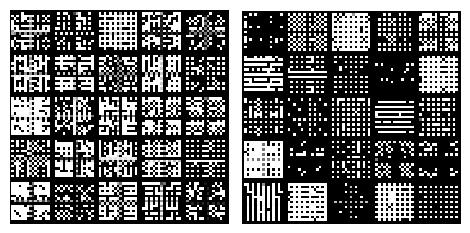}
        \caption{\label{fig:genft-sample}Sample of 25 synthetic feature maps from $ndf, ngf$ = 64 (left) and 512 (right).}
        \end{figure}

\begin{figure*}[h]
        \centering
        \includegraphics[width=1\textwidth]{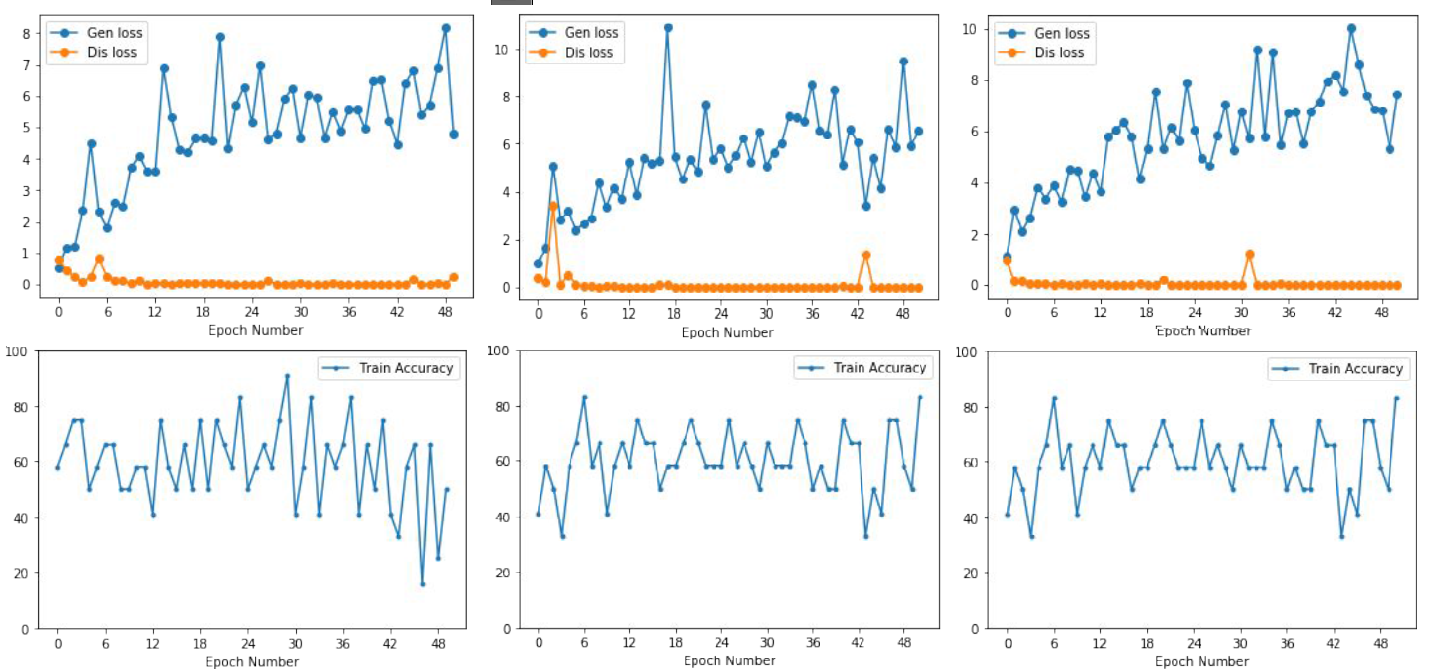}
        \caption{\label{fig:acgan-plots}Training loss (top) and classification accuracy (bottom) for ACGANs with $ndf, ngf$ = 64 (left), 256 (middle) and 512 (right).}
        \end{figure*}

        Qualitative evaluation of model outputs has low interpretability for feature maps as compared to images. Nevertheless, we show samples of first 25 synthetic feature maps (corresponding to the original features in Fig  \ref{fig:ft-sample}). It is evident from the larger number of grey and black areas that hidden layers of the model with large number of kernels (high-dim) retain more information in the generated features as shown in Fig.  \ref{fig:genft-sample}. Further, the training loss and discriminator accuracy (for class labels) curves show more stability in the high-dim setting as shown in Fig \ref{fig:acgan-plots}.
        

        After training, we generate 4800 samples with the ACGAN and carry out quantitative evaluation of the synthetic features using 2 measures:
        \begin{itemize}
            \item Classification accuracy of synthetic features given by $D$: this improves with increase in hidden layer size. The low-dim model ($ndf$ and $ngf$ = 64) results in 0.76 and the high-dim model ($ndf$ and $ngf$ = 512) results in a score of 0.823.
            
            \item Classification accuracy of the binary classifier on the test set (with original features only) when trained on original and and synthetic features.
        \end{itemize}

        
        \subsubsection{Model Final Results}
        The baseline binary-classifier is trained on original U-Net features. First, we trained a linear classifier on downsampled features. However, this model always predicts the HGG class (because of class imbalance). So, we concluded that downsampling using fully-connected (FC) layers results in information loss. Then, we use a CNN based binary classifier and experimented with different number of convolutional and FC layers. The final classifier is trained with two blocks of convolutional, batch normalization and max-pooling layers followed by two FC layers. We also use dropout ($p$= 0.5) before and after the first FC layer. All layers use relu activation and the last FC layer uses sigmoid for classification.  We obtained a classification accuracy of 61\%, on the test set. Note that 64\% of instances in the test set are from class HGG. This confirms that the U-Net features are not discriminative enough. 
        \newline \par
        Next, we augmented the dataset using the features generated by the ACGAN, for which the discriminator predicted the correct class label. We varied the proportion of real vs synthetic samples in the dataset and found that adding augmenting the dataset with upto 1200 synthetic features improves the per-class accuracy. However, adding more synthetic features hurts the model performance in that even though the overall accuracy increases, the model starts predicting mostly LGG or HGG only for the test sample. The best classification performance was achieved on a dataset augmented with 1158 synthetic features with 52.4\% overall accuracy (Table \ref{table:res}). Even though the overall score is lower compared to the baseline, there is a significant increase in the accuracy for the LGG class (from 5\% to 38.4\%) while the accuracy for the HGG class decreases. These results, together with the class-classification of the ACGAN, suggest that the ACGAN generates features that are more discriminative. This explains why the original and synthetic feature maps look different.
        
        \begin{table}[h!]
        \centering
        \begin{tabular}{|p{4.0cm}| p{1.0cm}| p{1.0cm}| p{1.0cm}|}
        \hline
         & HGG & LGG & Total \\
        \hline
        Baseline & 0.95 & 0.05 & 0.61\\ 
        Augmented Features Set & 0.569 &  0.384 & 0.524  \\  
        
        \hline
        \end{tabular} \\
        \caption{The total and per-class classification accuracy on the test set.}
        \label{table:res}
        \end{table}


\section{Conclusion and Future Work}

    In this work, we propose using U-net and ACGAN as a learning framework for feature generation of medical images followed by classification to validate the quality of generated features. Our experiments were based on the hypothesis that by using U-Net we can derive significant features as the U-Net is considered to be a state-of-the-art network for the purpose of medical image segmentation. We show promising results on the classification task by augmenting the dataset using synthetic features, generated from the ACGAN, with an increase in the accuracy of the LGG class from 5\% to 38.4\%. Although, in our experiment we only managed to collect and generate features from the last convolutional block of the U-net (5th block) due to dimensional and computational challenges, but we strongly recommend using features from the higher deconvolutional blocks of the U-Net with the skip connections, as theoretically, they are supposed to provide better feature representation of the images. 
    
    We use an ACGAN, where every generated sample has a corresponding class label, hence we use this architecture for using it as a class-conditioned feature generator. 
    For future experiments, we would like to use other GAN architectures eg.InfoGANs because they can be conditioned on latent embeddings of classes. Additionally, the model performance could be compared against a baseline ACGAN that generates synthetic images instead of features. Further, to overcome memory overflow issues due to high-dimensional features, images could be cropped using segmentation masks predicted by the U-Net, and then a CNN could be used to learn lower-dimensional features of the tumor regions.  


{\small
\bibliographystyle{ieee}
\bibliography{egbib}
}

\end{document}